\documentclass{article}

\usepackage{amsmath}
\usepackage{fullpage}
\usepackage{amsfonts}
\usepackage{amssymb}
\usepackage{amsthm}
\usepackage{array}
\usepackage{thmtools}
\usepackage{thm-restate}
\usepackage{mathtools}
\usepackage{bbm}
\usepackage{booktabs}
\usepackage{graphicx}
\usepackage[dvipsnames]{xcolor}
\usepackage{tikz}
\usepackage{nicefrac}
\usepackage{microtype}
\usepackage{wrapfig}
\usepackage{subcaption}
\usepackage[noend]{algorithmic}

\usepackage{url}
\usepackage{hyperref}
\hypersetup{
    colorlinks=true,
    linkcolor=blue,
    urlcolor=blue,
    citecolor = blue
}

\usepackage[capitalize,noabbrev]{cleveref}
\creflabelformat{equation}{#2#1#3}

\usepackage{fancyvrb}
\RecustomVerbatimEnvironment
  {Verbatim}{Verbatim}{numbers=left,frame=single,xleftmargin=2cm}

\newcommand{\sys}{Plume}
\newcommand{\naive}{Naive}
\newcommand{\speedy}{RuntimeOptimized}

\newcommand{\D}{\mathcal{D}}

\newcommand{\M}{\mathcal{M}}

\renewcommand{\c}{\boldsymbol{c}}

\newtheorem{definition}{Definition}

\newtheorem{lemma}{Lemma}

\begin{document}

\title{\sys: Differential Privacy at Scale}

\author{Kareem Amin \and Jennifer Gillenwater \and Matthew Joseph \and Alex Kulesza \and Sergei Vassilvitskii\\
\\
\texttt{\{kamin, jengi, mtjoseph, kulesza, sergeiv\}@google.com}\\
}
\date{}

\maketitle

\begin{abstract}
Differential privacy has become the standard for private data analysis, and an extensive literature now offers differentially private solutions to a wide variety of problems. However, translating these solutions into practical systems often requires confronting details that the literature ignores or abstracts away: users may contribute multiple records, the domain of possible records may be unknown, and the eventual system must scale to large volumes of data. Failure to carefully account for all three issues can severely impair a system's quality and usability.

We present \sys, a system built to address these problems. We describe a number of sometimes subtle implementation issues and offer practical solutions that, together, make an industrial-scale system for differentially private data analysis possible. \sys{} is currently deployed at Google and is routinely used to process datasets with trillions of records.
\end{abstract}

\section{Introduction}
\label{sec:introduction}

Differential privacy (DP) has become the standard for data anonymization. Since its introduction~\cite{DMNS06}, the vast majority of research has focused on developing new algorithms and mechanisms that improve the privacy-utility trade-off. However, in practice algorithm speed and ease-of-use can have just as much impact on the ultimate benefit of such a system. In this work we present \sys, a system for large-scale DP aggregations deployed at Google and used thousands of times per week, processing petabytes of data. 

The goal of differential privacy is to ensure that the attacker cannot meaningfully distinguish between outcomes where any particular user is present in or absent from the dataset, thereby protecting the user’s privacy. There has been extensive work developing algorithmic primitives that achieve this goal, leading to state-of-the-art mechanisms for private statistics, private machine learning, and private analytics. However, good mechanisms are just one piece of a production system. The overall system also needs to be fast, accurate, and easy to use. We faced three major challenges in achieving these goals:
\begin{enumerate}
    \item \textbf{The contribution bounding problem}: How should we limit the number of contributions from each user in the dataset?  This is a necessary step for bounding the sensitivity of DP mechanisms, but many existing systems address it in ways that result in ease-of-use and accuracy issues (see \cref{sec:related_work}).
    \item \textbf{The key selection problem}: How should we aggregate keyed data when the set of keys is not known in advance?  From a privacy and implementation standpoint, it is simpler to require system clients to provide a fixed, known set of keys, but this dramatically reduces ease-of-use.
    \item \textbf{The scalability problem}: How can we make high-accuracy solutions for these previous two problems scale to large datasets?
\end{enumerate}

As a running example, consider the hypothetical task of identifying popular recipe searches. Ignoring privacy considerations, this is a simple query over the set of observed searches: \texttt{GROUP BY} recipe and \texttt{COUNT} the number of queries for each. With privacy in mind, it is tempting to simply replicate the \texttt{GROUP BY} and replace the \texttt{COUNT} with a \texttt{PRIVATE COUNT}; however, two issues emerge. First, we need to determine the sensitivity of the count query---some users may have searched for a recipe once, while others may have searched for many different recipes before settling on one. Thus, an important part of the implementation is to bound the {\em contribution} of any user to the final outcome. This bounding introduces a bias to the results (due to dropped data), but it also reduces the variance of the noise required to achieve privacy. (See \cite{AKM19,EMM20} for more on this bias-variance trade-off.)

Second, which recipes can we even return counts for?  For instance, suppose that only one person searched for a particular misspelling of a recipe, e.g., ``kasagna'' instead of ``lasagna.''  Revealing the fact that this ``recipe'' had a non-zero count would also reveal the presence of that user in the dataset. More broadly, this is a question of handling unknown domains, where user records are not limited to a pre-defined set of keys. Determining which keys are safe to reveal is a process we refer to as {\em key selection}. In some applications the set of possible keys is fixed ahead of time (e.g., counting the number of searches from each country). However, in practice the set of keys often needs to be determined from the dataset, raising additional privacy concerns.

Finally, there is the challenge of doing all of this at scale. With multi-terabyte datasets being the norm, massively parallel systems building on the MapReduce framework are usually needed in practice. These systems scale to arbitrarily large datasets simply by introducing more worker machines. However, this comes at the cost of a restricted data flow; the computation proceeds in synchronous rounds during which each worker sees only a small fraction of the data, and reshuffling data across machines between rounds is expensive. This is a particular challenge for differential privacy, given the need to process data along multiple dimensions in order to bound user contributions, select safe keys, and aggregate the results.

\section{Related Work}
\label{sec:related_work}

Most DP systems can be categorized as working either in the ``local'' or the ``central'' model. In the local model, user data is anonymized before it is sent from a user-controlled device to a central server.  Examples of systems working in this model include~\cite{DKY17,EPK14,A17}.  In this work, we consider only the central model: sensitive user data has been gathered in a central location and a trusted curator applies a differentially private mechanism  to generate aggregations of the data.  Industrial uses of DP systems that operate in this model include Microsoft's PINQ~\cite{M09}, the FLEX~\cite{JNS18} system used by Uber, Google's DP SQL system~\cite{WZL20},  Google's Privacy on Beam system~\cite{GPB21}, and LinkedIn's Audience Engagement API~\cite{RSP21}.

PINQ~\cite{M09} and FLEX~\cite{JNS18} assume that each user only contributes a single record.  Unfortunately, for many real-world datasets, enforcing such an assumption often greatly reduces the dataset size and yields low-quality results; in our running example, this assumption would require pre-filtering the dataset of recipe searches to ensure that no user contributes more than one recipe search.  Handling multiple contributions per user is a large part of what makes our system work in practice.  More recent systems such as Google's DP SQL system~\cite{WZL20}, Google's Privacy on Beam system~\cite{GPB21}, and LinkedIn's Audience Engagement API~\cite{RSP21} also do not limit users to a single contribution.  We discuss these in more detail below.

Google's DP SQL system~\cite{WZL20} deals meaningfully with the three practical challenges discussed in \cref{sec:introduction}: (1) it allows users to contribute multiple records, (2) it handles situations where the domain of possible records is unknown, and (3) it does so in a way that scales.  However, it deals with (1) and (2) somewhat jointly, similar to what we will describe in \cref{sec:fast}.  This design choice makes the system very scalable, but can also reduce system accuracy; see \cref{sec:why_fast_is_bad} for details.  Google's Privacy on Beam system~\cite{GPB21} currently behaves similarly.

LinkedIn's Audience Engagement API~\cite{RSP21} also addresses these three practical challenges, making somewhat different trade-offs.  Similar to \cite{WZL20} and our work, LinkedIn's API handles (2), the unknown domain problem, by noising aggregates and thresholding to discard small values.  However, in order to handle (3), the scalability problem, it makes some sacrifices in how it handles (1), the multi-contribution problem.  In more detail: the API either assumes that users can contribute to all keys, or that the limit on the number of keys that they can contribute to has already been enforced on the input data.  In the former case, the API only returns results for the top-$k$ keys, in order to limit the sensitivity of the result.  In the latter case, the API suffers from the same accuracy issues as the system that we describe in \cref{sec:fast}.

Our work focuses on better handling of (1), the multi-contribution problem, at a relatively small cost to scalability.  Briefly, our system does contribution bounding as part of handling the key selection problem, just as is done in Google's DP SQL system~\cite{WZL20}, Google's Privacy on Beam system~\cite{GPB21}, and LinkedIn's Audience Engagement API~\cite{RSP21}. However, it then repeats contribution bounding on the original dataset once the domain has been fixed.  The main advantage of this is that user contributions are not wasted on keys that will not appear in the output.  Subsequent sections describe how to accomplish this in a manner that scales well.

Finally, we note that there are now a number of open-source differential privacy libraries~\cite{G21, HBML19, S20, O21}. However, none of these are end-to-end solutions for petabyte-scale data.
\section{Preliminaries}
\label{sec:setup}

At a high level, our goal is to apply an aggregation algorithm $A$ to the records associated with each key in a large dataset. $A$ is intentionally generic---it may, for instance, count the number of records associated with a key; in our recipe example, keying search records by their query text allows us to identify the most popular recipe queries. Alternatively, $A$ may compute the 99\textsuperscript{th} percentile of the values of the records; in a dataset keyed by app version this may be useful to understand changes to app latency, and so on. The system that we outline in this paper has been applied to compute many basic statistics such as counts, sums, means, variances, and quantiles for many diverse applications.

Formally, we assume the dataset $D$ consists of keyed records in $\D = K \times V$, where $K$ is the set of possible keys and $V$ is the set of possible values. Denote by $V_k(D)$ the collection of values that appear paired with key $k$ in $D$; our goal is to compute $A(V_k(D))$ for all keys $k \in K$ where $V_k(D)$ is nonempty. A user may contribute arbitrarily many records to the dataset; however, we will assume for now that a single value $v \in V$ is rich enough to describe all of a user's data for a particular key, and thus that a user contributes at most one record with each key. (This assumption is primarily to simplify the exposition, and we discuss how to lift it in \cref{sec:practical_considerations}.)

These types of simple keyed aggregations are easily achieved in the MapReduce model of distributed computation~\cite{DG04}, which implements large-scale algorithms using a collection of parallel worker machines. In this model, the input records are initially divided arbitrarily among the workers, and the computation then proceeds in rounds. In each round, a worker first independently processes each record in its local collection (the {\em map} phase), the processed records are then redistributed so that records with matching keys are located at the same workers (the {\em shuffle} phase), and finally the shuffled records associated with each key are processed as a group by their local workers (the {\em reduce} phase). For our recipe example, the map phase would key each input record by its query text, the shuffle phase would group the records on workers according to their keys, and finally the reduce phase would count the number of records for each query.

Of course, more complex algorithms can require more than one round. Since a shuffle requires moving large amounts of data across the relatively slow channels between workers, it is usually much more expensive than the map and reduce phases, which process records locally and in parallel. Therefore, the number of rounds is a good first-order predictor of a MapReduce algorithm's running time and has been identified as the key metric in MapReduce algorithm analysis \cite{KSV10}. Efficient implementations usually seek to keep the number of rounds small.

\subsection{Differential Privacy}

When working with datasets containing private user information, we must consider the privacy implications of the aggregations we compute. Differential privacy formalizes the extent to which aggregated values might still allow a knowledgeable attacker to learn information about a specific individual.

To make this precise, let $U$ denote the set of users that might contribute records to a dataset, and let $\mathbbm{1}_D(u)$ denote the indicator whose value is $1$ if user $u \in U$ contributes (one or more) records to $D$, and $0$ otherwise. We will call two datasets $D$ and $D'$ \emph{neighbors}, denoted $D \sim D'$, if they differ only in the presence of a single user: $|\{u \in U \mid \mathbbm{1}_D(u) \ne \mathbbm{1}_{D'}(u)\}| \leq 1$. Differential privacy limits an attacker's ability to distinguish between any two neighboring datasets, and thus to detect the inclusion of any single user (let alone learn the details of their data).

\begin{definition}[\textbf{Differential privacy}~\cite{DMNS06}]
A mechanism $M \colon \D \to Y$ is \emph{$(\epsilon, \delta)$-differentially private} if, for any two neighboring datasets $D \sim D'$ and any $S\subseteq Y$, we have
\begin{equation*}
    P(M(D) \in S) \leq e^\epsilon P(M(D') \in S) + \delta~.
\end{equation*}
\end{definition}

This definition guarantees that no attacker, regardless of prior knowledge or computational ability, can gain more than a fixed amount of information about any individual from the output of $M$. Generally, differentially private mechanisms must be stochastic; this ensures that neighboring datasets have a nonzero probability of producing the same output. The parameters $\epsilon$ and $\delta$ control the required amount of overlap between the output distributions: as $\epsilon$ and $\delta$ shrink toward zero, the distributions must be more similar, and hence more noise is required. At the same time, the privacy guarantee becomes stronger.

Note that the neighboring relation, which is key to the definition, is sometimes defined in terms of \emph{records} rather than users, with datasets said to be neighbors if they differ in the presence of a single record. This leads to a weaker notion of privacy in which an attacker gains limited information about any single record, but may be able to identify a user who contributes multiple records. In the real world, this is often unacceptable. As we are interested in building practical systems, we focus on so-called \emph{user-level} privacy instead\footnote{Note that our system can seamlessly handle privacy units other than users, simply by replacing the user IDs in the input dataset with the IDs of the alternative privacy unit.  For instance, to provide $(\epsilon, \delta)$-DP for groups of users instead of just individual users, user IDs can be replaced by group IDs.  The exact unit of privacy that is appropriate depends on the application.}. This stronger definition has significant implications for computation and efficiency, as we discuss in later sections.

\subsubsection{Properties}

Differential privacy has the useful property of \emph{composition}; that is, when the outputs of multiple private mechanisms are combined, the result is itself differentially private, with parameters that can be computed from those of the underlying mechanisms.

\begin{lemma}[\textbf{Basic composition}~\cite{DMNS06}]
\label{lem:comp}
    Let $M_1, \ldots, \M_r$ be $r$ mechanisms that respectively satisfy $(\epsilon_1, \delta_1)$-$, \ldots, (\epsilon_r, \delta_r)$-differential privacy. Then the combined output of $M_1, \ldots, M_r$ satisfies $\left(\sum_{i=1}^r \epsilon_i, \sum_{i=1}^r \delta_i\right)$-differential privacy overall.
\end{lemma}

More advanced general composition results are also known (e.g., \cite{KOV15}). Additionally, some mechanisms may support tighter composition results that depend on the specifics of their design (e.g., the exponential mechanism's non-adaptive composition guarantee~\cite{DDR20}). We will assume basic composition here for simplicity, but our construction also supports these more advanced results when they are applicable.

A second useful property of differential privacy that we rely on is \emph{post-processing}. Simply stated, applying any public function to the output of a differentially private mechanism does not affect its privacy guarantees. (Here ``public'' only means that the function itself does not depend on private information.)

Taken together, the composition and post-processing properties allow us to run multiple private mechanisms and then combine and manipulate their results while maintaining an overall privacy guarantee.

\section{Mechanism}

Returning to our original goal, while in principle we would like to exactly apply the aggregation algorithm $A$ to the records associated with each key in the dataset, in practice we will only aim to get as close to this as possible \emph{while remaining $(\epsilon, \delta)$-differentially private}. Although there are many different ways to approach this problem, we will develop a simple, structured mechanism that covers a wide array of practical use cases and fits well in the MapReduce framework.

Specifically, we will assume that a differentially private algorithm $M$ has been provided as a surrogate for $A$. Our goal will be to apply $M$ to as many of the records associated with as many of the keys as possible, subject to the limits of privacy that we describe below. We will assume that $M$ can be efficiently executed for an arbitrary choice of parameters $(\epsilon_M, \delta_M)$ during the reduce phase of a MapReduce round\footnote{Depending on the MapReduce implementation and the choice of $M$, it might be possible to execute the mechanism in various ways. For instance, if $M$ can be decomposed into an associative operation with pre/post-processing steps, which is common, then it might be possible to run the reduction hierarchically using multiple workers to save time. We will not focus on these details here, but they can also be important in achieving an efficient, practical result.}, but otherwise the details of $M$ are not especially important: $M$ could be the Laplace mechanism \cite{DMNS06}, an exponential mechanism \cite{MT07}, or any other differentially private mechanism that can be run on each key in parallel.

\subsection{Contribution Bounding}

Following \cref{lem:comp}, applying $M$ with parameters $(\epsilon_M, \delta_M)$ to $r$ different keys yields a composed privacy guarantee of $(r\epsilon_M, r\delta_M)$. This implies that the mechanism parameters should be chosen as roughly $\epsilon_M = \epsilon / r$, $\delta_M = \delta / r$ to obtain an overall $(\epsilon, \delta)$-differentially private result.

While this is a possible approach, for many realistic settings $r$ will be prohibitively large and $\epsilon_M$ and $\delta_M$ will be small. The results will consequently be very noisy. We therefore enforce the additional constraint that no individual user contributes records for more than $L$ different keys, where $L$ is an input parameter that can be adjusted to the application. Under this assumption, neighboring datasets $D \sim D'$ can have differing output distributions on only at most $L$ unique keys. The resulting privacy is thus equivalent to composing $M$ just $L$ times, making the overall privacy guarantee $(L\epsilon_M, L\delta_M)$. This means that we need only scale our privacy parameters by $L$, regardless of the total number of keys.

Of course, in a real-world dataset, a user might in fact contribute records to many different keys (perhaps they are an avid cook, or a poor typist). Thus, we must actually \emph{enforce} our assumption within the system in order to have an easy-to-use, end-to-end system that achieves private results. This step, which we refer to as contribution bounding, is typically assumed away in the differential privacy literature. In practice enforcing contribution bounds on large, distributed datasets can be computationally difficult. This is one of the major challenges we address in later sections.

The contribution bounding algorithm we will use is (conceptually) straightforward. For each user $u \in U$, let $K^{(u)}$ denote the set of keys for which $u$ contributes a record. We will select $\min(L, |K^{(u)}|)$ elements of $K^{(u)}$ uniformly at random, and then discard any of $u$'s contributions to unselected keys. This approach guarantees that our assumption holds without introducing any unnecessary bias.\footnote{In general, $M$ will also have its own contribution bounding requirements; for instance, the Laplace mechanism requires bounded sensitivity so that the appropriate amount of noise can be determined. We assume such requirements are enforced as part of the execution of $M$, e.g., by first clamping the user values to the allowed range. See \cref{sec:practical_considerations} for more discussion.}\footnote{We could replace this selection algorithm with a weighted version, as in the baseline algorithms from \cite{GGSSY20}, without any changes to the overall structure of the system described here. However, there does not seem to be any straightforward way to make the more complex ``Policy Laplace'' and ``Policy Gaussian'' methods from that work scalable.}

\subsection{Key Selection}
\label{sec:key_selection}

Contribution bounding removes our dependence on the \emph{number} of keys. However, the question still remains: which keys, exactly, should we apply $M$ to? In general, the set of all possible keys $K$ will be too large to enumerate; in the recipe example $K$ is the set of all possible recipe queries, i.e., all strings. Moreover, ignoring the problem of size, keys that do not appear in the dataset would only pollute the result with noise. At the same time, we cannot simply return a result for every key that appears in the dataset, since a key (such as ``kasagna'') might reveal the presence of a specific user. We must therefore ensure that the set of keys appearing in the output map is itself differentially private.

To achieve this, we will rely on existing techniques for private set selection, such as the thresholding on Laplace-noised unique user counts first proposed by \cite{KKM09}. Specifically, we will count the number of unique users contributing to each key (considering only those keys that actually appear in the dataset), and then call a stochastic decision function \texttt{DP\_RETAIN\_KEY} to determine whether or not each count is sufficient to retain the associated key. In general, keys with larger counts will be retained with higher probability, but the only strict requirement on \texttt{DP\_RETAIN\_KEY} is that the resulting set of keys, assuming each user initially contributes to at most $L$ different keys, must be differentially private.\footnote{Note that we have now used the $L$ assumption twice: once for composing the applications of $M$, and once for scaling our key selection mechanism. We will look at some subtle implications of this in \cref{sec:fast}.} We will refer to the collection of selected keys as $S$, with $(\epsilon_S, \delta_S)$ denoting the privacy parameters of the selection process.

\subsection{Privacy}

We can now summarize the privacy-impacting steps of our overall mechanism. All steps operate under the assumption that our system has restricted each user's contributions to at most $L$ keys. We first select a set of keys $S$ for which an aggregation will be produced; this step is $(\epsilon_S, \delta_S)$-differentially private. We then apply $M$ to the records associated with each selected key and return the results; this step is $(L\epsilon_M, L\delta_M)$-differentially private. Finally, applying post-processing and composition, we conclude that our overall mechanism is $(\epsilon_S + L\epsilon_M, \delta_S + L\delta_M)$-differentially private. We assume that the parameters have been chosen in advance such that this guarantee meets the needs of the application, i.e., that $\epsilon = \epsilon_S + L\epsilon_M$ and $\delta = \delta_S + L\delta_M$.
\section{Naive Implementation}\label{sec:naive}

Some of the challenges in implementing our mechanism are already apparent. For instance, contribution bounding requires the dataset to be grouped by user, whereas key selection requires grouping by key. The shuffles needed to switch between these views can be very expensive. Thus, we should design the implementation carefully. For simplicity, we begin with a straightforward approach. This approach will be deliberately inefficient, but it gives us a starting point for discussing various optimizations. 

Given the preliminaries from the previous section, we conduct the following five operations sequentially:
\begin{enumerate}
\item perform contribution bounding to limit the number of keys associated with any single user to $L$, producing the bounded dataset $D_L$,
\item apply a private key-selection mechanism to $D_L$ to generate a safe key set $S$,
\item restrict the original dataset $D$ to the keys in $S$, producing $D_S$,
\item bound the contributions of users in $D_S$ to $L$ keys each, producing a dataset $D_\mathrm{safe}$, and finally
\item apply the mechanism $M$ to each key in $D_\mathrm{safe}$.
\end{enumerate}

Some of these stages might appear computationally wasteful. For example, (1) and (4) appear to be doing the same work. As we will see in later sections, incorrect optimizations of these redundancies will come at a cost to utility.  

\textbf{Step 1 --- Contribution Bounding:} We first describe how to limit each user to $L$ random keys in a MapReduce framework. We assume the existence of an associative data structure \texttt{Heap<Key, Value>(User user, int key\_limit)}, where we will set $\texttt{key\_limit} = L$. A call to \texttt{Heap::Insert(Key key, Value value)} inserts the given \texttt{(key, value)} pair into the heap.  Position within the heap is determined by \texttt{key}. The ordering over keys is determined by a random hash function unique to the user that seeded the heap. In other words, each user is assigned a different random order over keys. Furthermore, the heap only retains data associated with the top \texttt{key\_limit} keys in the heap. Standard \texttt{Heap} operations (insertion, lookup, merge) can therefore be implemented in $O(L \log L)$ time and $O(L)$ memory in the worst-case. In our first application, we do not need to associate keys with values, and therefore use \texttt{Heap<Key>} to denote the analagous container where \texttt{Value} is a null type. Employing this \texttt{Heap}, a map-reduce for bounding user contributions is now straightforward.

\begin{Verbatim}
contribution_bound_map(User u, Key k):
    Heap(u, L) heap
    heap.Insert(k)
    emit (u, heap)
\end{Verbatim}
\begin{Verbatim}
contribution_bound_reduce(User u, List<Heap<Key>> keys):
    Heap(u, L) result
    for heap in keys:
      result.Merge(heap)
    emit result  
\end{Verbatim}

After \texttt{contribution\_bound\_reduce} is complete, each heap contains a random selection of at most $L$ keys contributed by a single user. 

\textbf{Step 2 --- Key Selection:} It is now straightforward to count the occurrences of each key using another map-reduce, as in the pseudocode below.  Any number of DP set-selection strategies can be executed on the resulting counts, including simply adding noise and applying a threshold as described in \cref{sec:key_selection}. (More advanced techniques, such as \cite{DVG21}, could also be used.) In the code below, \texttt{DP\_RETAIN\_KEY} is an arbitrary key selection strategy.

\begin{Verbatim}
heap_to_key_map(Heap<Key> heap):
    for key in heap.Top(L):
      emit (key, 1)
\end{Verbatim}
\begin{Verbatim}
key_occurrences_reduce(Key key, List<int> counts):
    sum = 0
    for count in counts:
      sum = sum + count
    emit (key, sum)
\end{Verbatim}
\begin{Verbatim}
apply_privacy_map(Pair<Key, int> key_count):
   if DP_RETAIN_KEY(key_count.count):
     emit key_count.key
\end{Verbatim}

\textbf{Step 3 --- Joining with Selected Keys:} After the previous step, we have a differentially private set of keys $S$. Next, we generate $D_S$ by joining the original data $D$ with with the key set $S$, filtering so that only data associated with the keys of $S$ is retained. We consider a simple reduce-side join. This will generate a major inefficiency that we discuss in greater detail in section \ref{sec:ineff}. In short, this join will force one reducer to process all the data associated with a single key in $S$. 

\begin{Verbatim}
join_result_reduce(Key key, boolean key_in_S,
                   List<Pair<User, Value>> data_from_D):
    if not key_in_S:
      return
    for (user, value) in data:
      emit(user, key, value) 
\end{Verbatim}

\textbf{Steps 4 and 5 --- Aggregation:} In order to complete the computation of $D_\mathrm{safe}$, we must once again restrict $D_S$ to contain $L$ elements. This proceeds analogously to the contribution bounding described in the previous section. \verb|contribution_bound_map| and \verb|contribution_bound_reduce| are applied to the resulting data, replacing instances of \texttt{Heap<Key>} with \texttt{Heap<Key, Value>}. After these stages, we have restricted the data in $D$ to keys in $S$ and limited the contribution of each user to $L$ different keys.

At this point, it is possible, due to the randomness of the heaps, that some key from $S$ is associated with no data in $D_\mathrm{safe}$.  However, for the overall system output to be differentially private, we must output a value for each of the keys in the selected set $S$.  To ensure that this happens, we add to $D_\mathrm{safe}$ a special dummy value for each key in $S$.  This is not an efficiency bottleneck, as it is a simple union operation combining $D_\mathrm{safe}$ and a dataset consisting of a single key-value pair for each key in $S$.

Finally, we can apply the aggregation mechanism $M$ to the result.

\begin{Verbatim}
aggregation_map(Heap<Key, Value> heap):
    for (key, value) in heap.Top(L):
      emit (key, value)
\end{Verbatim}

\begin{Verbatim}
aggregation_reduce(Key key, List<Value> key_data):
    emit (key, M(key_data))
\end{Verbatim}

\subsection{Inefficiencies}\label{sec:ineff}
\begin{figure}[h]
\centering
\includegraphics[scale=.09,trim={0 8cm 0cm 0},clip]{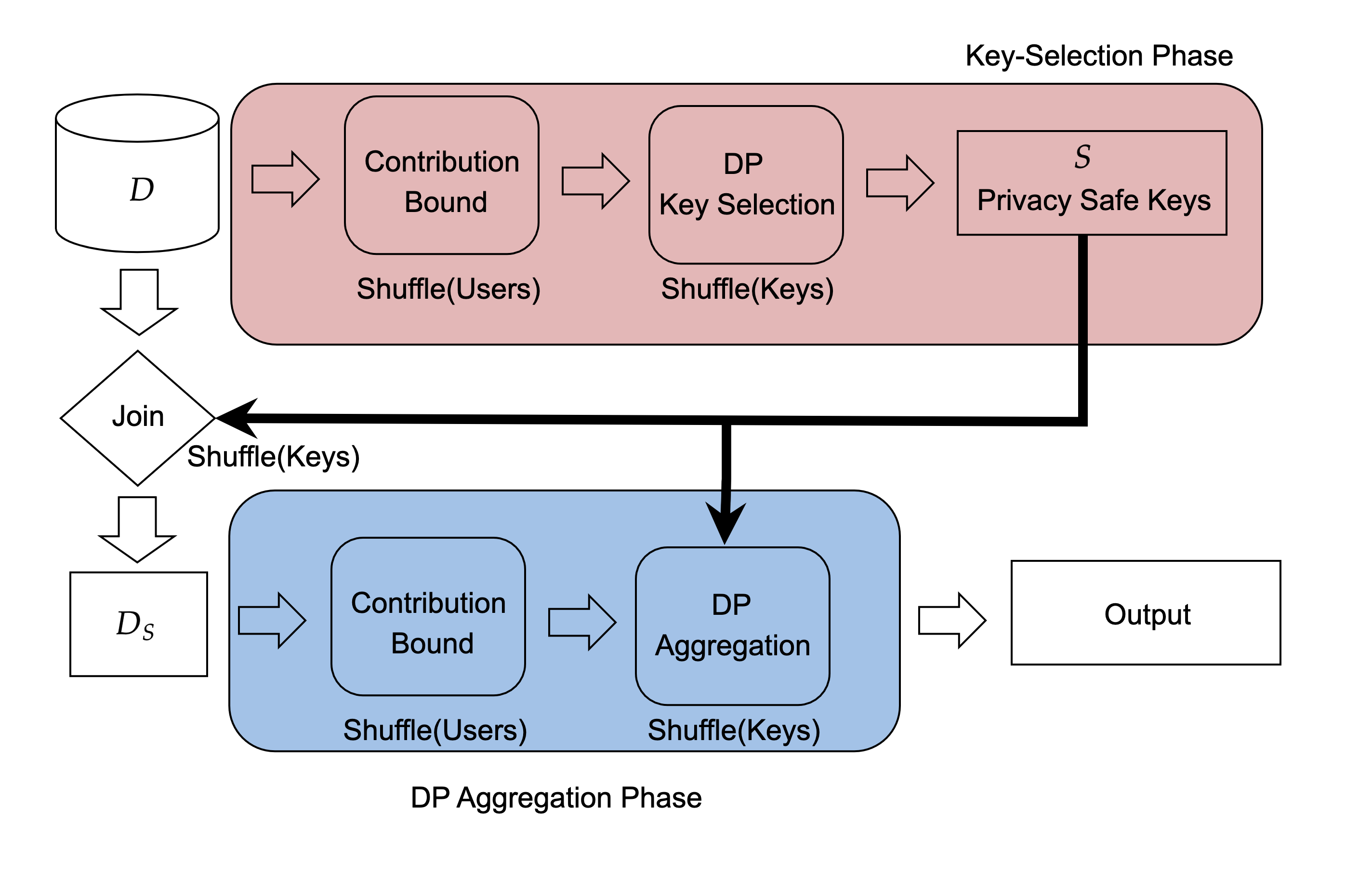}
\caption{Naive DP Aggregation}
\label{fig:naive}
\end{figure}

Putting together the procedures from the previous sections gives us the ingredients for an end-to-end system for executing generic DP aggregation queries, summarized in~\cref{fig:naive}.

While such a system produces DP results, it will not do so very efficiently. First, the communication complexity of shuffle stages tends to be large in any parallel computation framework. Here, we utilize five different shuffles to arrive at the DP result. Moreover,~\cref{fig:naive} suggests a parallel between the key selection phase of the pipeline and the DP aggregation phase of the the pipeline; they both consist of contribution bounding, followed by a shuffle on keys and application of a DP subroutine. Is it possible to parallelize these phases, or even do the work using the same shuffles?  Finally, there is the matter of the join. On real data, this join often represents a substantial bottleneck, because it is frequently very skewed. For instance, consider input data that has a long tail of low-count keys, many of which will not make it into the set of privacy-safe keys $S$.  This kind of situation is very common, and in fact applies to our running example of recipe search query counts.  In such a situation, the set $S$ is much smaller than the data $D$ that it's being joined with.  Such a skewed reduce-side join can be very inefficient.

\section{An (Over) Optimization}\label{sec:fast}

In light of the computational pitfalls discussed in the previous section, we seek an optimized version of the same system. Given the apparent parallel structure of the pipeline depicted in \cref{fig:naive}, it is tempting to merge the key-selection phase with the DP aggregation phase. Looking even more holistically, the pipeline consists of two shuffles on users and three shuffles on keys. It is tempting to arrange the work so that we need only two shuffles -- one on users and one on keys. This is indeed possible, and comes with the additional benefit of eliminating the costly join. 

The solution is as follows. We keep \textbf{Step 1} from \cref{sec:naive}, generating the dataset $D_L$. We then parallelize all the remaining work of the naive pipeline around a single shuffle on keys, essentially merging \textbf{Step 2, Step 3}, and \textbf{Step 5} of the naive solution. 

\begin{Verbatim}
aggregation_map(Heap<Key, Value> heap):
    for (key, value) in heap.Top(L):
      emit (key, 1, value)
\end{Verbatim}

\begin{Verbatim}
aggregation_reduce(Key key, List<int> key_counts, List<Value> key_data):
    if DP_RETAIN_KEY(key_counts.sum):
        emit (key, M(key_data))
\end{Verbatim}

This drastically reduces the communication complexity of the pipeline, restricting all the work to two shuffles, with corresponding reduce stages that can likely be implemented as associative combines (depending on $M$). Moreover, the costly imbalanced join of the naive solution is completely eliminated. Unfortunately, this comes at a cost to utility. 

\begin{figure}[h]
\label{fig:fast}
\centering
\includegraphics[scale=.09,trim={0 4cm 10cm 0},clip]{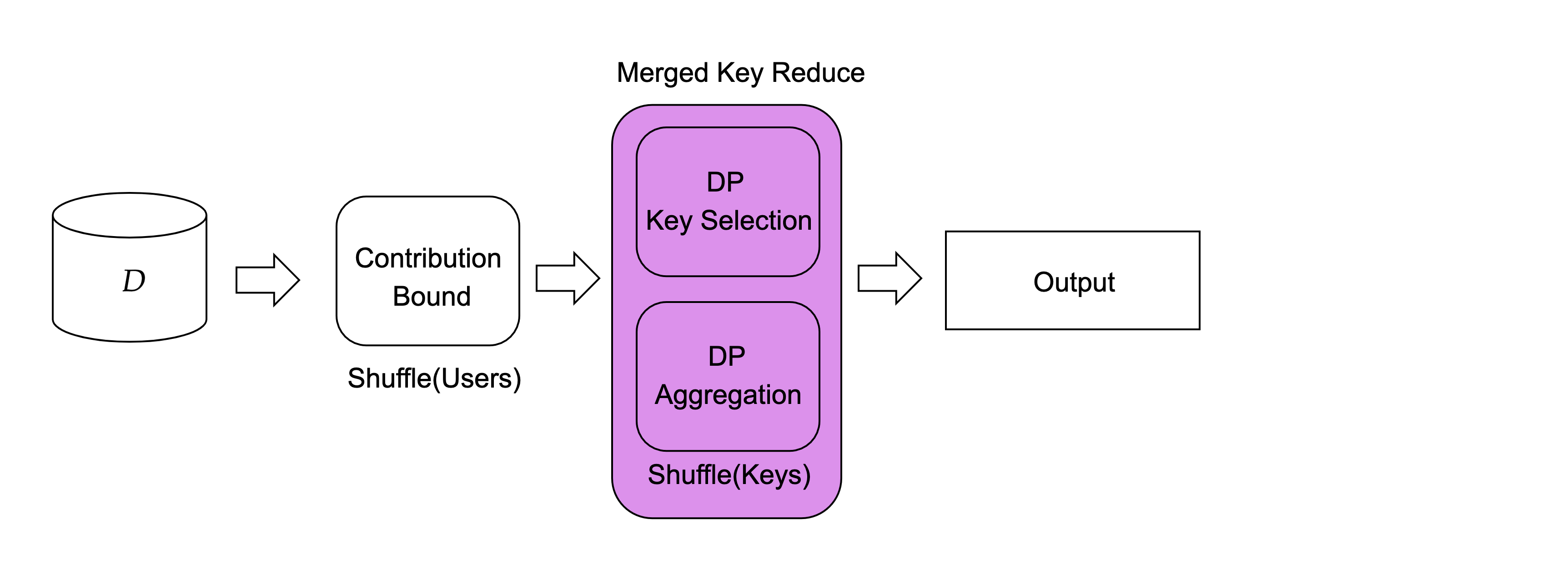}
\caption{Runtime Optimized DP Aggregation}
\end{figure}

\subsection{Utility Reduction}
\label{sec:why_fast_is_bad}

While the two solutions sketched above might appear equivalent at first, they are in fact not. In the first, naive solution, contribution bounding happens twice: initially, we compute $D_L$ as a precursor to deriving a safe set of keys $S$. We then compute our final aggregate on $D_\mathrm{safe}$ by bounding the contributions in $D_S$. In our zeal to optimize, the second solution only executes the restriction operation once, computing $D_L$. This dataset is used (in parallel) to compute $S$, as in the naive solution, but also to filter on $S$. Ultimately this results in executing $M$ on $D_L$ restricted to the key set $S$, rather than $D_\mathrm{safe}$, which is the restriction of $D_S$ to satisfy the contribution bound $L$. Below, we give simple count query example that illustrates how contribution bounding twice can have a significant positive impact on the accuracy of system outputs.

Consider a toy dataset where all users visit two locations: a home location, which is unique to them, and a popular landmark (Times Square, the Eiffel Tower, etc). We wish to count visits to locations. If the bound $L$ is 1, a single round of contribution bounding will under-count Times Square by a factor of 2. With enough users, the under-counted statistic is still sufficient for Times Square to survive key selection. 

However, no matter how many users we have, there is a 2x relative error in the count associated with Times Square, and all the popular landmarks. Using a second round of contribution bounding, as in the naive solution, saves us. In this second round, we randomly select one location for each user but, crucially, this selection is from the keys that have survived key selection. For the hypothetical location dataset, this procedure essentially amounts to selecting the landmark associated with each user.  

We note that this situation is not as contrived as it might seem in the above toy example. The same effect manifests whenever the frequency of keys in the data follow a long-tail distribution, with each user contributing keys that are relatively common and keys that are relatively rare.

\section{Plume: Best of Both Worlds}\label{sec:plume}

Is it possible to retain most of the computational speed-ups of \cref{sec:fast} without taking a utility hit relative to the solution from \cref{sec:naive}? In this section we describe how to thread the needle between both solutions, retaining the utility characteristics of the naive solution, while only requiring one additional shuffle beyond that used for the computationally-optimized solution. 

We begin with the same shuffles as the naive solution: the first on $D$, shuffled by user in order to compute $D_L$, and the second on $D_L$, shuffled by key in order to compute the privacy-safe key set $S$. At this point we diverge from the naive solution, which executes a reduce-side join between $S$ and the initial dataset $D$  (see \cref{fig:naive}). As already discussed, this can be a bottleneck if not handled carefully, as $S$ will often contain a small fraction of the keys from $D$.

It is tempting, therefore, to replace this reduce-side join with a map-side join between $S$ and $D$. There are two concerns with such an approach. First, there is no guarantee that $S$ will be small enough to fit in memory in order to perform a standard map-side join. Secondly, subsequent to the join, the data $D_S$ will still need to be shuffled by users in order to apply the second round of contribution-bounding. This repeats work that was already executed to shuffle $D$.

Since the cardinality of $S$ is relatively cheap to compute with a single parallelizable reduce, we can leverage this to overcome the first concern, tailoring our join to the characteristics of $S$. If $|S|$ is small, copying $S$ within each mapper is the most straightforward and efficient way to proceed. For larger $|S|$, we make use of a read-only distributed hash table~\cite{BDE+20,KLM+14}, replacing the memory requirements of a map-side reduce with the I/O overhead of performing lookups into a DHT. At the same time, we do not join $S$ with $D$ directly. Instead, we make use of the fact that we have already shuffled $D$ on $U$, and have at our disposal the result of this shuffle, $D_0$, typed as $(U, \mathrm{List}\langle K, V\rangle)$. Let $T$ denote the lookup table whose implementation depends on $|S|$. We pass $T$ to a mapper on $D_0$, and can now apply a second round of contribution bounding, this time filtering only for records in $T$, generating $D_\mathrm{safe}$.

Note that, as in \cref{sec:naive}, it is again possible at this point that $D_\mathrm{safe}$ does not contain values for some keys in $S$.  As before, we add to $D_\mathrm{safe}$ a special dummy value for each key in $S$ to ensure that there is at least one value associated with each. Thus, there will be a DP aggregation result for each key of $S$.

There is a potential inefficiency here in that a single worker might have to evaluate every record in $\mathrm{List}\langle K, V\rangle$. 
In practice, it makes little difference. Even on petabyte-sized datasets, the number of records associated with a single \emph{user} tends to be manageable by a single worker. 

The final stage of our approach shuffles $D_\mathrm{safe}$ on keys, applying $M$ as in the naive approach of ~\cref{sec:naive}.

\begin{figure}[t!]
\label{fig:final}
\centering
\includegraphics[scale=.08999]{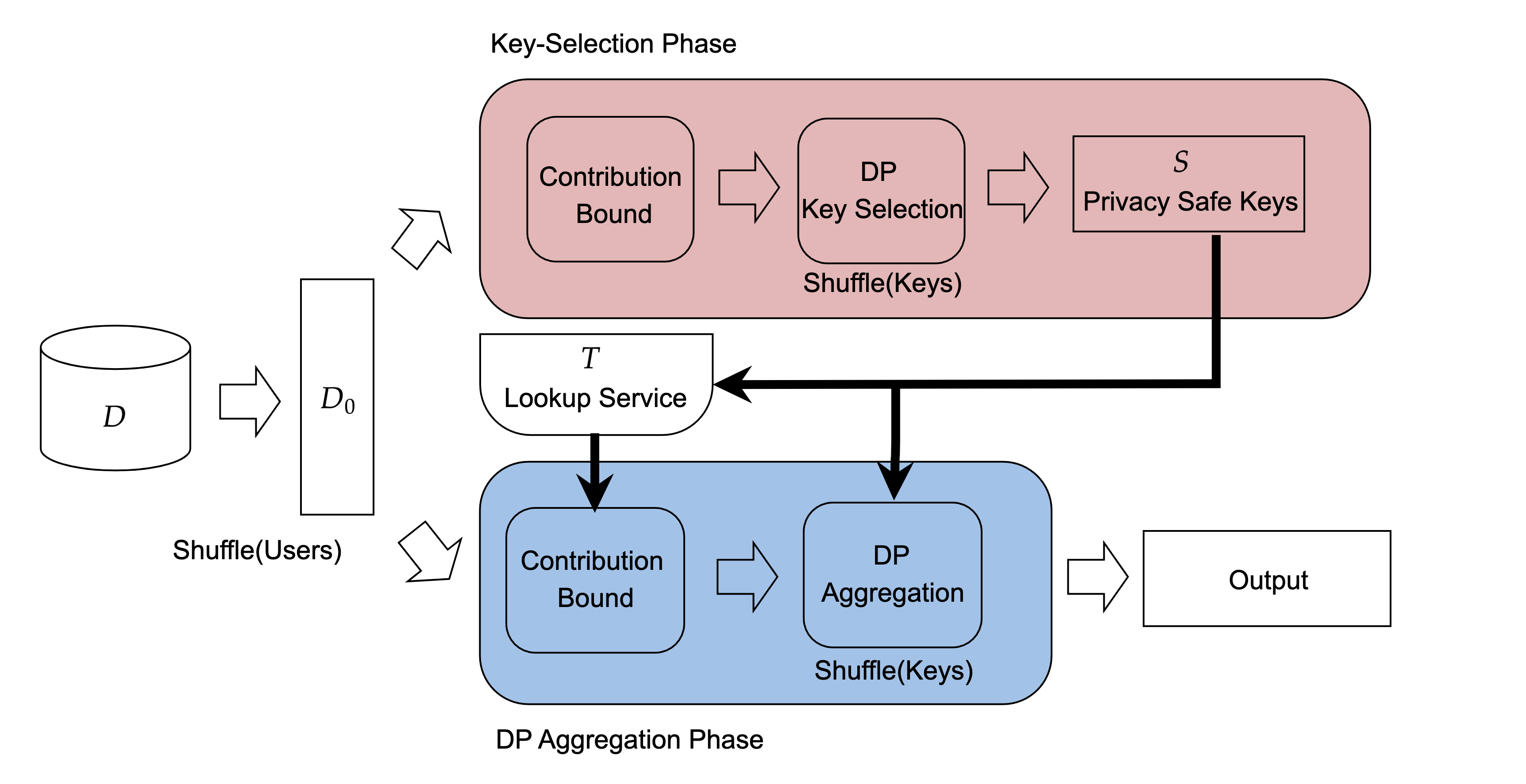}
\caption{Plume DP Aggregation}
\end{figure}

\section{Practical Considerations}
\label{sec:practical_considerations}

We now discuss number of additional practical considerations. So far we have thought about $M$ as a black box that handles its own DP aggregation. In practice, it is often beneficial to expose some of the internal workings of $M$ to the larger execution framework in order to take advantage of several additional optimizations.  

For instance, we have assumed that the mechanism $M$ can enforce its own contribution bounds during the reduce. In practice, however, contribution bounding requires collating records by user, and a user's data may not be stored contiguously, making this inefficient. Moreover, even if a user's records will be simultaneously accessible, it still might be a good idea to combine them into a more succinct representation earlier in the pipeline, after Plume has executed its first group by on users. This reduces the dataset size and thus communication complexity for all downstream stages. 

Consider, for example, a dataset storing users' engagement on a website in seconds. There might be multiple values, represented as floating-point numbers, corresponding to a single user and key, and passing all of these values throughout the various stages of Plume will inflate I/O costs, probably unnecessarily. If, for example, the mechanism is computing a DP sum, then it is semantically appropriate to reduce all values corresponding to a user and key to a single number representing all the total number of seconds the user spent on the website. Note that such a reduction is specific to the mechanism $M$. The same reduction might not be semantically appropriate for a different mechanism (e.g., quantiles).

Another issue that we have glossed over is the practical benefit of implementing reducers as associative combiners. This allows the system to take full advantage of parallelism and reduce the communication costs of the computation even further. Without any knowledge of the internal workings of $M$, the system must treat $M$ as an arbitrary reducer, forgoing any of these optimizations.

Consider again the example of computing a differentially private sum. For each key, $M$ receives a collection of floating-point numbers corresponding to individual measurements, and outputs a DP sum as a result. $M$ implemented as a Laplace mechanism might first clip each value to some endogenously defined maximum value $L_\infty$, then sum each of these clipped values, and then finally add appropriate noise to satisfy the DP guarantee. If this is visible to the larger system, then the clipping can be mapped as a preprocessing stage after the first shuffle on users. This frees us up to implement the addition step as hierarchical combine on the last reduce, followed by a final map to add the privacy noise. 

In practice, our system turns these observations into requirements. We enforce that $M$ is more than a black-box reducer, but can be composed by these four operations: a combiner on raw inputs, a preprocessing map, a combiner on keyed data, and a final noiser. Each of the mechanisms supported by our system adhere to this decomposition allowing us to further optimize performance. 

\section{Experiments}
\label{sec:experiments}

To demonstrate the usability and scalability of our solution, and to validate the analysis above, we provide performance measurements using both synthetic and real-world datasets. We evaluate three systems: $\mathrm{\naive{}}$, $\mathrm{\speedy{}}$, and $\mathrm{\sys{}}$, which correspond to Sections \ref{sec:naive}, \ref{sec:fast}, and \ref{sec:plume}, respectively. Our aggregations will be simple counting tasks, described in more detail below.

We report system runtime alongside two measures of aggregation error, each of which compares the system's output for each key with the exact (non-private) result. The first measure is absolute error, which is the mean (per retained key) absolute difference between the system result and the exact result. The second measure is relative error, which is the mean (per retained key) of the absolute error divided by the exact result. In this case we always omit keys dropped during selection.

For all experiments, we fix  $\epsilon = \ln(3)$ and $\delta = 10^{-5}$ as overall privacy parameters\footnote{These privacy parameters are arbitrary, and the exact values are not particularly important for illustrating the differences between the systems that we compare.  Meaningful privacy guarantees may require tighter parameters in practice.}, and then set the privacy parameters for the key selection phase to $\epsilon_S = \epsilon/2$, $\delta_s = \delta$. The privacy parameters for the mechanism $M$ are consequently $\epsilon_M = \epsilon / (2L)$, $\delta_M = 0$. (None of the mechanisms we use in these experiments require $\delta_M$.) The parameter $L$ is chosen based on the properties of the dataset; see \cref{sec:tuning} for a more detailed discussion on setting this parameter.

\subsection{Synthetic Data}\label{sec:synthetic_experiments}

Our synthetic datasets are designed to capture the heavy-tailed nature of real data, but can be easily generated at arbitrary scale simply by adding more synthetic users. To generate the dataset, each user draws their number of records i.i.d. from a distribution with range $[1, 10^5]$ and mean $10$. The distribution is heavy-tailed (Zipf-Mandelbrot) and its parameters are chosen so that there is a roughly $\nicefrac{1}{3}$ probability of generating more than $10$ contributions\footnote{Specifically, the probability of sampling $x$ is proportional to $(x + q)^{-s}$ where $s = 4.67$ and $q = 25$.}. 
Each contributed record has a key sampled i.i.d. from a set of size $10^6$, again using a heavy-tailed distribution, where the first $10^3$ keys have a total probability of about $\nicefrac{1}{3}$.\footnote{Zipf-Mandelbrot with parameters s = 1.4, q = 1000.} We denote the synthetic dataset containing $n$ distinct users as $D_\textrm{synth}[n]$, and our experiments include datasets for $n \in \{10^5, 10^6, 10^7, 10^8, 10^9\}$. The uncompressed size of $D_\textrm{synth}[10^9]$ is $171$GB, containing roughly $10^{10}$ records.

For these datasets, our target aggregation algorithm $A$ simply counts the number of records associated with each key, and for our differentially private $M$ we use the Laplace mechanism on these counts. $M$ applies its own contribution bound to enforce finite sensitivity; specifically, each user is allowed to contribute at most $1$ record to each key. $M$ then adds Laplace noise with scale parameter $1 / \epsilon_M$ (see \cite{DMNS06} for details).  We set the contribution bound $L = 64$.

All performance measures are averaged across $5$ runs, with shaded regions displaying standard error. The key selection phase is identical across the systems (modulo the exact values of noise that happen to be drawn), and the results in \cref{fig:key_selection} confirm that the numbers of keys retained by each approach are statistically identical. Thus, we only report error metrics for retained keys.

\begin{table}
\begin{center}
\begin{tabular}{||c l l l l||} 
 \hline 
  &&&&\\ [-0.9em]
  & $10^6$ Users & $10^7$ Users & $10^8$ Users & $10^9$ Users \\  
  & ($178$ MB) & ($1.8$ GB) & ($18$ GB) & ($171$ GB) \\ [0.5ex] 
 \hline\hline
 \naive{} & $934\ (\pm 11)$ & $8994\ (\pm 19)$ & $50722\ (\pm 44)$ & $257892\ (\pm 47)$ \\ 
 \hline
 \speedy{} & $936\ (\pm 8)$ &  $9015\ (\pm 4)$ & $50737\ (\pm 45)$ & $257895\ (\pm 107)$ \\
 \hline
 \sys{} &  $939\ (\pm 6)$ & $8992\ (\pm 12)$ & $50738\ (\pm 58)$ & $257963\ (\pm 60)$ \\
 \hline
\end{tabular}
\end{center}
\caption{Average number of keys retained on synthetic data, standard devation is shown in parenthesis.}
\label{fig:key_selection}
\end{table}

\begin{figure}[ht]
    \centering
    \includegraphics[scale=0.5]{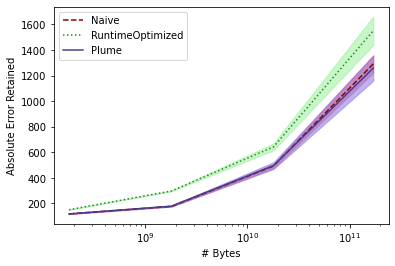}
    \includegraphics[scale=0.5]{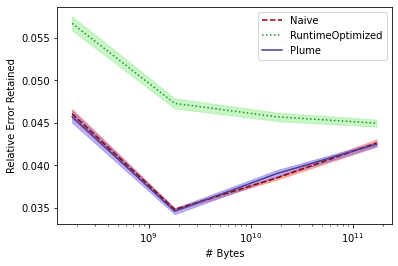}
\caption{Aggregation errors on synthetic data}
\label{fig:error_count}
\end{figure}

\cref{fig:error_count} shows the aggregation error produced by each system for different dataset sizes. There is a significant gap between \naive{}/\sys{} and \speedy{} due to the second contribution bounding phase, with \naive{} and \sys{} performing significantly better since they do not waste contributions on keys that have already been dropped. 

We can understand the relationship between error and dataset size by noting that aggregation error comes primarily from two sources: contribution bounding and noise. Since we use a fixed value of $L$ and users are i.i.d., the absolute error due to contribution bounding tends to grow with the number of users (more data is being bounded away), while the noise remains constant. Thus, overall, measures of absolute error per key grow with the number of users. By the same token, the relative error due to contribution bounding is (in the limit) constant, while the relative error due to noise decreases as counts grow, thus relative error converges toward a constant value. (Prior to convergence the set of selected keys is also changing, which can result in increasing error due to contribution bounding; this explains the non-monotonic behavior in \cref{fig:error_count}.)

\begin{figure}[ht]
    \centering
    \includegraphics[scale=0.5]{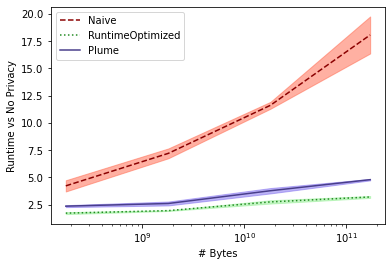}
\caption{Slowdown vs no privacy on synthetic data}
\label{fig:runtime_count}
\end{figure}

In \cref{fig:runtime_count} we report runtime in relative terms, dividing the wall time required for a given system by the wall time required to execute the aggregation without any privacy constraints.  For all systems, runtime increases with dataset size. The relative slowdown for \naive{} increases dramatically as the size of the datasets increases. In contrast, \speedy{} and \sys{} have smaller penalties of at most $3$x and $4$x, respectively, for even the largest datasets.
 
Overall, then, there is (as the name suggests) a clear runtime advantage for \speedy{} compared with \naive{}, which is up to seven times slower on the largest datasets. However, the aggregation error of \speedy{} is significantly higher. \sys{} retains the accuracy of \naive{} while preserving the majority of speed-up attained by \speedy{}.

\subsection{Reddit Data}

We next apply these systems to real-world data. We use the popular webis-tldr-17-corpus, which consists of 3.8 million posts associated with users on the content and discussion website Reddit~\cite{SVPS17}. %
Our task is to generate frequency counts for each word in the corpus. Words correspond to keys in the database, and a record associates a key/word with an integral value identifying the number of times a particular user uttered the word. We would like to sum the utterance counts for each word across all users.

Our differentially private mechanism $M$ will again be the Laplace mechanism, now applied to the sums of utterance counts for each key/word. $M$ enforces its own contribution bound by allowing each user to contribute at most $8$ utterances to each word, and then adds Laplace noise with scale parameter $8 / \epsilon_M$. We set the contribution bound $L = 512$.

\cref{fig:reddit} summarizes the results. As in the synthetic experiments, we report runtime as a factor of the wall time required to execute the aggregation without privacy constraints. Once again, all methods execute key selection identically, so our error metrics only consider retained keys. Results are averaged over $15$ runs with standard errors reported in parentheses. 

\begin{table}
\begin{center}
\begin{tabular}{||c l l l||} 
 \hline 
  &&& \\[-0.9em]
   & Runtime Factor  & Absolute (Retained) & Relative (Retained) \\ [0.5ex] 
 \hline\hline
 \naive{} & $13.26\ (\pm 1.05)$ & $113{,}214\ (\pm 6{,}663)$ & $\mathbf{0.198\ (\pm 0.001)}$ \\ 
 \hline
 \speedy{} & \ $\mathbf{2.63\ (\pm 0.42)}$  & $128{,}550\ (\pm 6{,}617)$ &  $0.273\ (\pm 0.001)$ \\
 \hline
 \sys{} & \ $\mathbf{2.76\ (\pm 0.29)}$  & $\mathbf{101{,}038\ (\pm 5{,}400)}$ & $\mathbf{0.196\  (\pm 0.002)}$ \\
 \hline
\end{tabular}
\end{center}
\caption{Performance on Reddit data with standard error shown in parenthesis. Values minimizing each column (up to standard error) are depicted in bold.}
\label{fig:reddit}
\end{table}

Once again, \sys{} allows us to achieve the best of both worlds. \sys{} attains error that is comparable to \naive{}, while \speedy{} has significantly worse accuracy, with approximately $38\%$ higher relative error on retained keys. Conversely, the running time of \sys{} is indistinguishable from \speedy{}, while \naive{} is significantly slower, requiring roughly $5$x longer to run.

\subsection{Parameter Tuning}\label{sec:tuning}

\begin{figure}[ht]
    \centering
\begin{subfigure}{0.45\textwidth}
\centering
    \includegraphics[scale=0.5]{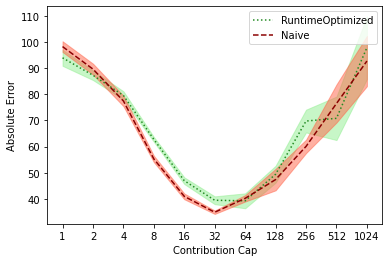}
\end{subfigure}
\begin{subfigure}{0.45\textwidth}
\centering
    \includegraphics[scale=0.5]{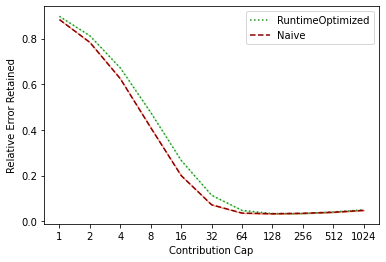}
\end{subfigure}
\caption{\label{fig:caps_synthetic}Error vs the contribution bound $L$ on synthetic data}
\end{figure}

In practice, tuning the contribution bound $L$ can be critical to achieving good performance. If $L$ is too small, much of the data may be discarded, leading to poor results. On the other hand, if $L$ is too large, then an excessive amount of noise may be required. While in the experiments above we used fixed values of $L$, here we investigate the relationship between $L$ and aggregation error in more detail. 

\cref{fig:caps_synthetic} shows how the choice of $L$ affects the synthetic count results on dataset $D_\textrm{synth}[10^7]$. Since the choice of $L$ affects the key selection process, on the left we show the mean absolute error over \emph{all} keys, where dropped keys are treated as having an aggregated value of $0$. We plot relative error on retained keys on the right.

First, note that the advantages of \sys{} hold across the board; thus, even if in practice a more sophisticated parameter tuning approach is used, \sys{} will still reliably give the best results. Second, though, it is apparent that the choice of $L$ can have a much larger impact than the choice among the systems studied here. Thus, the ability to efficiently bound contributions under arbitrary $L$ is a significant practical advantage for \emph{any} differentially private system, compared to those that support only $L=1$ or assume that bounding has been performed in advance (see \cref{sec:related_work}). For instance, on this dataset, a naive choice of $L = 1$ would have produced roughly $2.4$ times the absolute error, and $19$ times the relative error. 

We note in passing that, although for our experiments we used plots like these to fix roughly optimal values of $L$, this approach is not itself differentially private since it relies on a comparison with the exact aggregations. In practice, other selection methods should be used to preserve user privacy (see discussion in \cref{sec:conclusion}).

\section{Conclusion \& Future Work}
\label{sec:conclusion}

In this work we presented a practical system for differential private aggregation that handles the three challenges outlined in \cref{sec:introduction}: the contribution bounding problem, the key selection problem, and the scalability problem. \sys{} has been used at Google for several years by a multitude of teams to formally anonymize their results.

Many interesting challenges remain. First, as evidenced in our experiments, tuning user contribution limits often has a large impact on accuracy. Integrating the tuning of these parameters into the differential privacy system would greatly improve usability. As the tuning must itself be done in a differentially private manner for the system to be end-to-end differentially private, this presents interesting challenges~\cite{AKM19,EMM20,PS22}. Another area for future work is the further improvement of privacy-utility trade-offs. Here, even small constant factor improvements, such as judiciously selecting among Laplace, Gaussian, truncated geometric~\cite{DVG21}, or other noise types, can have a large impact on practical usability. Yet another direction is tackling scalability problems in higher order primitives, expanding the number of scalable differentially private algorithms. Finally, while \sys{} is a batch system, there is significant interest in differentially private systems that process streaming data in an incremental fashion.
\section*{Acknowledgments}

We would like to thank Per Anderson, Christoph Dibak, Miguel Guevara, Andrés Muñoz Medina, and Jane Shapiro for their critical work in making \sys{} possible. In addition, we would like to thank members of Google's core anonymization team for their contributions to \sys: Mirac Vuslat Basaran, Pern Hui Chia, Damien Desfontaines, Vadym Doroshenko, Alain Forget, Bryant Gipson, Dennis Kraft, Sasha Kulankhina, Milinda Perera, Daniel Simmons-Marengo, Yurii Sushko, and Xinyu Ye.

\bibliographystyle{plain}
\bibliography{references}

\end{document}